\documentclass{article}%
\usepackage{amssymb}%
\usepackage{amsmath}%
\usepackage{amsfonts}%
\usepackage{graphicx}
\providecommand{\U}[1]{\protect\rule{.1in}{.1in}}

\begin{document}

\title{Exact closed equation for reduced equilibrium \\distribution functions of the many-particle system }
\author{Victor F. Los
\and Institute of Magnetism, Nat. Acad. of Sci. of Ukraine,
\and 36-b Vernadsky Blvd., 03142 Kiev, Ukraine}
\maketitle

\begin{abstract}
An exact closed equation for $s$ - particle equilibrium distribution function
($s<N$) of the system of $N\gg1$ interacting particles is obtained. This
integra-differential $\beta$ - convolution equation ($\beta=1/k_{B}T$) follows
from the Bloch equation for the canonical distribution function by applying
the projection operator integrating off the coordinates of $N-s$ irrelevant
particles. The method of expansion of the obtained equation kernel in the
particle density $n$ is suggested. The solution to this equation in the linear
in $n$ approximation for the kernel is found.

\end{abstract}

\section{\bigskip Introduction}

In equilibrium statistical mechanics, calculation of thermodynamic functions
can be performed either with the help of the corresponding partition function
or by means of the reduced distribution functions. For many-body nonideal
systems, both approaches are rather involved and based on the Ursell-Mayer
cluster expansion of the partition function and distribution functions (see,
e.g. \cite{Isihara 1971} and references herein). The particle density
expansions of the equation of state and of a pair distribution function for a
nonideal gas can then be obtained. However, for dense gases and liquids these
density expansions poorly converge or even diverge.

This problem is conventionally cured with the help of the equations for the
reduced distribution functions. For example, one may use the BBGKY hierarchy
of coupled integra-differential (in coordinate variables) equations for the
reduced distribution functions and close it by employing some approximation
procedure, e.g., the Kirkwood superposition approximation (see, e.g.
\cite{Bogoliubov}). This approximation cannot however be strictly justified.
There are other approaches leading to the approximate closed equations for a
pair distribution function such as the Percus-Yevick (PY) nonlinear integral
equation and the Hypernetted Chain (HNC) equation (see, e.g. \cite{Balescu}).
The approximations used for obtaining these equations can be analyzed in terms
of the Mayer diagrams (the PY and HNC approximations are correct to the first
order diagram in the particle density) but, again, these approximations are
rather guessed than substantiated.

The principal and interesting question then arises: Is it possible to derive
an \textbf{exact} closed equation for $s$-particle distribution function
($s<N$) for the system of $N$ ($N\gg1$) interacting particles? The goal of
this paper is to demonstrate that, surprisingly, this is possible. We will
derive such closed equation by applying the projection operator technique to
the Bloch differential equation (with respect to $\beta=1/k_{B}T$) for the $N$
- particle canonical classical Gibbs distribution function. The obtained exact
equation is the integra-differential equation for the $s$-particle
distribution function with a complicated integral ($\beta$ - convolution)
kernel. This equation can be approximately solved by, e.g., the expansion of
the kernel in the particle density series. The expansion of the kernel,
however, is much more effective than the expansion of the distribution
function itself. For example, the linear in the particle density term of the
kernel expansion corresponds to the summing up of the infinite series of the
distribution function density expansion.

\section{Projection operator derivation of the exact closed equation for a
reduced distribution function}

We start with the classical equilibrium Gibbs distribution function for the
system of $N\gg1$ interacting classical particles with coordinates
$x_{i}=\{x_{i}^{\alpha}\}$ ($i=1,\ldots,N,\alpha=1,2,3$) occupying a volume
$V$ at temperature $T$
\begin{align}
D_{N}(\beta)  & =D_{N}(\beta;\{x_{i}\}_{N})=\frac{1}{Z_{N}(\beta)}\exp(-\beta
U_{N}),\beta=1/k_{B}T,\nonumber\\
Z_{N}(\beta)  & =\int dx_{1}\ldots\int dx_{N}\exp(-\beta U_{N}),\nonumber\\
\{x_{i}\}_{N}  & =x_{1},\ldots,x_{N}\label{1}%
\end{align}
where $U_{N}$ is the system's potential energy depending on the whole set of
spatial variables $\{x_{i}\}_{N}$
\begin{equation}
U_{N}=\sum\limits_{1\leq i<j\leq N}\Phi_{ij},\Phi_{ij}=\Phi(\left\vert
x_{i}-x_{j}\right\vert )\label{2}%
\end{equation}
and integration is performed over the whole system's volume $V$.

Let us consider the quantity
\begin{equation}
f_{N}(\beta)=f_{N}(\beta;\{x_{i}\}_{N})=Z_{N}(\beta)D_{N}(\beta)=\exp(-\beta
U_{N}).\label{2a}%
\end{equation}
The Bloch equation for function (\ref{2a}) reads%
\begin{equation}
\frac{\partial f_{N}(\beta)}{\partial\beta}=-U_{N}f_{N}(\beta).\label{3}%
\end{equation}
The formal solution to this equation is
\begin{equation}
f_{N}(\beta)=\exp[-(\beta-\beta_{0})U_{N}]f_{N}(\beta_{0}).\label{3a}%
\end{equation}
If we choose $\beta_{0}=0$, the initial value of the distribution function
(\ref{2a}) $f_{N}(\beta_{0})$ is
\begin{equation}
f_{N}(0)=1.\label{3b}%
\end{equation}

Let us introduce the operator%
\begin{equation}
P_{s}=\frac{1}{V^{N-s}}\int dx_{s+1}\ldots\int dx_{N},s<N\label{4}%
\end{equation}
acting on any function in the coordinate space, where integrations are
supposed to be performed over the whole system's volume $V$. Evidently, this
operator is a projection one satisfying the condition $P_{s}P_{s}=P_{s}$. The
complementary operator $Q_{s}=1-P_{s}$ is also a projection operator.

Now we apply the operators $P_{s}$ and $Q_{s}$ to Eq. (\ref{3}). This
procedure is completely equivalent to the well known approach leading to the
generalized master equations for the relevant distribution functions in the
kinetic theory (see, \cite{Nakajima}, \cite{Zwanzig}, \cite{Prigogine}). The
only difference is that our formalism is related to the temperature domain
instead of the time domain in the kinetic theory. Thus we have
\begin{align}
\frac{\partial f_{s}^{r}(\beta)}{\partial\beta}  & =-P_{s}U_{N}[f_{s}%
^{r}(\beta)+f_{N}^{i}(\beta)],\nonumber\\
\frac{\partial f_{N}^{i}(\beta)}{\partial\beta}  & =-Q_{s}U_{N}[f_{N}%
^{i}(\beta)+f_{s}^{r}(\beta)].\label{5}%
\end{align}
The relevant distribution $f_{s}^{r}(\beta)$ in our case is%
\begin{align}
f_{s}^{r}(\beta)  & =f_{s}^{r}(\beta;\{x_{i}\}_{s})=P_{s}f_{N}(\beta
)=\frac{Z_{N}(\beta)}{V^{N}}F_{s}(\beta),\nonumber\\
\{x_{i}\}_{s}  & =x_{1},\ldots,x_{s},\label{6}%
\end{align}
where $F_{s}(\beta)$ is the $s$-particle distribution function depending on
the set $\{x_{i}\}_{s}$ of the coordinates of $s<N$ particles (see
\cite{Bogoliubov})
\begin{equation}
F_{s}(\beta)=F_{s}(\beta;\{x_{i}\}_{s})=V^{s}\int dx_{s+1}\ldots\int
dx_{N}D_{N}(\beta),\label{6a}%
\end{equation}
which is normalized in such a way that
\begin{equation}
\frac{1}{V^{s}}F_{s}(\beta)dx_{1}\ldots dx_{s}\label{6b}%
\end{equation}
is the probability for finding the particles of the given group of $s$
particles in the infinitesimal volumes $dx_{1},\ldots,dx_{s}$ near the points
with the coordinates $x_{1},\ldots,x_{s}$. The irrelevant distribution
function is then given by%
\begin{align}
f_{N}^{i}(\beta)  & =f_{N}^{i}(\beta;\{x_{i}\}_{N})=Q_{s}f_{N}(\beta
)\nonumber\\
& =f_{N}(\beta)-\frac{1}{V^{N-s}}\int dx_{s+1}\ldots\int dx_{N}f_{N}%
(\beta).\label{7}%
\end{align}
Note, that the irrelevant distribution function (\ref{7}) depends on the
complete set of variables $\{x_{i}\}_{N}$, while the relevant function
(\ref{6}) is the reduced distribution function of interest depending on the
coordinates $\{x_{i}\}_{s}$ of the cluster of $s<N$ particles. The reduced
distribution functions (\ref{6a}) are sufficient for calculations of the
thermodynamic functions as the averages of the corresponding dynamical
functions which depend on the much smaller number of variables than the
original distribution function $D_{N}(\beta)$. Especially important is the
pair distribution function $F_{2}(\beta;x_{1},x_{2})$ which enables obtaining
the system's equation of state.

Now we need to solve the system of two equations (\ref{5}). The solution of
the second equation (\ref{5}) for the irrelevant distribution function reads
\begin{align}
f_{N}^{i}(\beta)  & =%
{\textstyle\int\limits_{\beta_{0}}^{\beta}}
d\beta_{1}\exp[-Q_{s}U_{N}(\beta-\beta_{1})](-Q_{s}U_{N})f_{s}^{r}(\beta
_{1})\nonumber\\
& +\exp[-Q_{s}U_{N}(\beta-\beta_{0})]f_{N}^{i}(\beta_{0}),\label{9}%
\end{align}
Substituting (\ref{9}) into the first equation (\ref{5}), we arrive at the
integra-differential equation for the relevant distribution function
$f_{s}^{r}(\beta)$, with a source $-P_{s}U_{N}\exp[-Q_{s}U_{N}(\beta-\beta
_{0})]f_{N}^{i}(\beta_{0})$ containing the irrelevant part of the distribution
function $f_{N}^{i}(\beta_{0})$ depending on the whole set of variables
$x_{1},\ldots,x_{N}$. In the kinetic theory this source represents the initial
(at the initial moment of time $t_{0}$) correlations in the many-particle
system and poses some problem to deal with (see \cite{Los}). Fortunately, in
our case, if we choose the initial value of the temperature parameter
$\beta_{0}=0$ in (\ref{9}), then, as it is seen from (\ref{3b}) and (\ref{7}),%
\begin{equation}
f_{N}^{i}(0)=0.\label{9a}%
\end{equation}

Thus, we obtain the following exact closed equation for the $s$-particle
distribution function $f_{s}^{r}(\beta)$%
\begin{align}
\frac{\partial f_{s}^{r}(\beta)}{\partial\beta}  & =-P_{s}U_{N}f_{s}^{r}%
(\beta)\nonumber\\
& +P_{s}U_{N}%
{\textstyle\int\limits_{0}^{\beta}}
d\beta_{1}\exp[-Q_{s}U_{N}(\beta-\beta_{1})]Q_{s}U_{N}f_{s}^{r}(\beta
_{1})\}.\label{10}%
\end{align}
Equation (\ref{10}) can be further specialized. It is useful to split the
energy $U_{N}$ (\ref{2}) into two parts%
\begin{align}
U_{N}  & =U_{s}+U_{s,N-s},\nonumber\\
U_{s}  & =\sum\limits_{1\leq i<j\leq s}\Phi_{ij},U_{s,N-s}=\sum\limits_{i=1}%
^{s}\sum\limits_{j=s+1}^{N}\Phi_{ij}+\sum\limits_{s+1\leq i<j\leq N}\Phi
_{ij}.\label{11}%
\end{align}
Here $U_{s}$ is the energy of the cluster of $s$ interacting particles with
coordinates $x_{1},\ldots,x_{s}$ ($s$ - cluster), the first term in
$U_{s,N-s}$ is the potential energy of interaction of $s$ - cluster with
remaining $N-s$ particles, and the second term in $U_{s,N-s}$ is the
interaction energy of the of $N-s$ particles not belonging to the $s$ -
cluster. Note, that the projector $Q_{s}$ commutes with $U_{s}$, $Q_{s}%
U_{s}=U_{s}Q_{s}$ (the same is of course true for $P_{s}$), but it is not the
case for $Q_{s}$($P_{s}$) and $U_{s,N-s}$. Thus, $\exp[-Q_{s}U_{N}(\beta
-\beta_{1})]=\exp[-(U_{s}Q_{s}+Q_{s}U_{s,N-s})(\beta-\beta_{1})]$.

For what follows, the expansion of the operator exponent
\begin{align}
e^{(A+B)\beta}  & =e^{A\beta}+\int\limits_{0}^{\beta}d\beta_{1}e^{A(\beta
-\beta_{1})}Be^{(A+B)\beta_{1}}\nonumber\\
& =e^{A\beta}+\int\limits_{0}^{\beta}d\beta_{1}e^{A(\beta-\beta_{1}%
)}Be^{A\beta_{1}}\nonumber\\
& +\int\limits_{0}^{\beta}d\beta_{1}\int\limits_{0}^{\beta_{1}}d\beta
_{2}e^{A(\beta-\beta_{1})}Be^{A(\beta_{1}-\beta_{2})}Be^{A\beta_{2}}%
+\ldots\label{12}%
\end{align}
is useful, where $A$ and $B$ are generally noncommuting operators.

Then, accounting for (\ref{12}) and the properties of projection operators
($P_{s}Q_{s}=0$,$P_{s}^{2}=P_{s}$, $Q_{s}^{2}=Q_{s}$), Eq. (\ref{10}) can be
represented as%
\begin{align}
\frac{\partial f_{s}^{r}(\beta)}{\partial\beta}  & =-(U_{s}+P_{s}%
U_{s,N-s})f_{s}^{r}(\beta)\nonumber\\
& +P_{s}U_{s,N-s}\int\limits_{0}^{\beta}d\beta_{1}e^{-Q_{s}U_{s,N-s}\beta_{1}%
}Q_{s}U_{s,N-s}e^{-U_{s}\beta_{1}}f_{s}^{r}(\beta-\beta_{1}).\label{13}%
\end{align}
This exact closed equation for the $s$ -particle function $f_{s}^{r}(\beta)$
can also be rewritten as%
\begin{align}
\frac{\partial f_{s}^{r}(\beta)}{\partial\beta}  & =-(U_{s}+P_{s}%
U_{s,N-s})f_{s}^{r}(\beta)\nonumber\\
& -P_{s}U_{s,N-s}\int\limits_{0}^{\beta}d\beta_{1}\frac{\partial}%
{\partial\beta_{1}}(e^{-Q_{s}U_{s,N-s}\beta_{1}})e^{-U_{s}\beta_{1}}f_{s}%
^{r}(\beta-\beta_{1}).\label{14}%
\end{align}
Note, that the operator $Q_{s}U_{s,N-s}\beta_{1}$ commutes with $U_{s}%
\beta_{1}$.

Equation (\ref{10}), or rather its more explicit versions (\ref{13}) and
(\ref{14}), is our main result which demonstrates the existence of the exact
closed equation for the reduced $s$ - particle distribution function. The
known closed equations, which follow from the decoupling of the BBGKY chain or
such as the Percus-Yevick (PY) and the Hypernetted Chain (HNC) nonlinear
integral equations for the pair distribution function, are approximate
equations in the coordinate space. The obtained closed equation is the exact
one but in the space-temperature domain. The process, described by this
equation, may be regarded as the evolution of the distribution function for
$s$ - cluster with particles coordinates $\{x_{i}\}_{s}$ at infinite
temperature ($\beta=0$) to the same configuration at the heat bath temperature
$T=1/k_{B}\beta$. This equation for the $s$ -particle function $f_{s}%
^{r}(\beta)$ (\ref{6}), differing from the conventional $s$ - particle
distribution function (\ref{6a}) only by the temperature dependent factor
$Z_{N}(\beta)/V^{N}$, is the complicated integra-differential non-Markovian
($\beta$ - convolution) equation. However, the existence of such an equation
is of principal importance and rather surprising.

\section{Solution of the equation for a reduced distribution function in the
Markovian approximation}

We should offer some recipe for resolving the obtained equation. One of the
possibilities is the expansion of the kernel of equation (\ref{13}) or
(\ref{14}), e.g. in the particle density $n=N/V$. Such an expansion is, as a
rule, much more effective than the direct expansion of the distribution
function itself. Each term of such an expansion generally accounts for the
partial infinite series of terms (diagrams) in the expansion of the
distribution function. Thus, the sums of these infinite series of terms may
obey the closed equation which approximates the exact equation (\ref{13}).
This fact is important for finding the reduced distribution function (e.g. the
pair distribution function) in the dense liquids and gases.

The operator exponent $e^{-Q_{s}U_{s,N-s}\beta_{1}}$in Eqs. (\ref{13}) and
(\ref{14}) can be expanded in series either directly in the powers of
$Q_{s}U_{s,N-s}\beta_{1}$ or (by using (\ref{12})) in the powers of
$P_{s}U_{s,N-s}$ as%
\begin{align}
e^{-Q_{s}U_{s,N-s}\beta_{1}}  & =e^{-U_{s,N-s}\beta_{1}}+\int\limits_{0}%
^{\beta_{1}}d\beta_{2}e^{-U_{s,N-s}(\beta_{1}-\beta_{2})}P_{s}U_{s,N-s}%
e^{-U_{s,N-s}\beta_{2}}\nonumber\\
& +\int\limits_{0}^{\beta_{1}}d\beta_{2}e^{-U_{s,N-s}(\beta_{1}-\beta_{2}%
)}P_{s}U_{s,N-s}\nonumber\\
& \times\int\limits_{0}^{\beta_{2}}d\beta_{3}e^{-U_{s,N-s}(\beta_{2}-\beta
_{3})}P_{s}U_{s,N-s}e^{-U_{s,N-s}\beta_{3}}+\ldots.\label{15}%
\end{align}

It is seen from (\ref{14}) and (\ref{15}) that the Eq. (\ref{14}) can be
expanded in the series in $P_{s}U_{s,N-s}$. In the zero approximation in
$P_{s}U_{s,N-s}$, Eq. (\ref{14}) (or (\ref{13})) can be easily solved and the
solution is
\begin{equation}
f_{s}^{r}(\beta)=\exp(-U_{s}\beta),\label{16}%
\end{equation}
which represents the distribution function for the isolated $s$ - cluster.
Equation (\ref{16}) coincides with the known result for the correlation
function $F_{s}(\beta;x_{1},\ldots,x_{s})$ (\ref{6}) (see, e.g.
\cite{Bogoliubov}) in the zero approximation in $n$ (an isolated $s$-
cluster), because in this approximation
\begin{align}
\frac{Z_{N}(\beta)}{V^{N}}  & =V^{-s}\int dx_{1}\ldots\int dx_{s}%
\prod\limits_{1\leq i<j\leq s}(1+f_{ij})\nonumber\\
& =1+V^{-1}\frac{s(s-1)}{2}\int dx_{2}f_{12}\nonumber\\
& +V^{-2}\frac{s(s-1)(s-2)}{3!}\int dx_{2}\int dx_{3}f_{12}f_{13}%
+\ldots,\nonumber\\
f_{ij}  & =\exp(-\Phi_{ij}\beta)-1,\label{16'}%
\end{align}
where $f_{ij}$ is the Mayer function, and we take into account that $f_{ij}$
depends on the difference of coordinates $\left\vert x_{i}-x_{j}\right\vert $.
Now, taking the limit $V\rightarrow\infty$ (while $s$ remains finite), we see
that in this limit the normalized partition function (\ref{16'}) $Z_{N}%
(\beta)/V^{N}=1$ and, therefore, $F_{s}(\beta)=$ $f_{s}^{r}(\beta)=\exp
(-U_{s}\beta)$ in the zero approximation in $n$ ($P_{s}U_{s,N-s}$). The terms
in Eqs. (\ref{13}), (\ref{14}), proportional to $P_{s}U_{s,N-s}$, describe the
interaction of $s$ - cluster with the "environment" of the remaining $N-s$ particles.

Now, we will restrict ourselves to the first approximation in $P_{s}U_{s,N-s}
$ of the Eq. (\ref{14}) kernel. We will show now, that each term containing
$P_{s}U_{s,N-s}$ is proportional to at least the first power of $n$. Using
(\ref{4}) and (\ref{11}), we have
\begin{align}
P_{s}U_{s,N-s}  & =P_{s}\sum\limits_{i=1}^{s}\sum\limits_{j=s+1}^{N}\Phi
_{ij}+P_{s}\sum\limits_{s+1\leq j<k\leq N}\Phi_{jk}\nonumber\\
& =\frac{V^{N-s-1}}{V^{N-s}}(N-s)\sum\limits_{i=1}^{s}\int dx_{s+1}%
\Phi(\left\vert x_{i}-x_{s+1}\right\vert )\nonumber\\
& +\frac{1}{2}\frac{V^{N-s-2}}{V^{N-s}}(N-s)(N-s-1)\int dx_{s+1}\int
dx_{s+2}\Phi(\left\vert x_{s+1}-x_{s+2}\right\vert ).\label{16''}%
\end{align}
For a many-particle system with $N\gg1$ (and $N\gg s$), Eq. (\ref{16''}) takes
the form%
\begin{equation}
P_{s}U_{s,N-s}=n\sum\limits_{i=1}^{s}\int dx_{s+1}\Phi_{is+1}+\frac{n^{2}}%
{2}\int dx_{s+1}\int dx_{s+2}\Phi_{s+1,s+2}.\label{16'''}%
\end{equation}
Although, the second term in the right-hand side of (\ref{16'''}) is formally
of the second order in the density $n$, it is additionally proportional to the
system volume $V$ due to the fact that $\Phi_{s+1,s+2}$ depends on the
difference $\left\vert x_{s+1}-x_{s+2}\right\vert $ of coordinates. Thus,
strictly speaking, the terms of all orders in $\Phi_{ij}$ ($s+1\leq i<j\leq
N$) should be accounted for. On the other hand, these terms do not depend on
the particles' coordinates and, therefore, contribute only to the temperature
dependent factor.

Because the integrand of Eq. (\ref{14}) is at least of the first order in
$P_{s}U_{s,N-s}$, we can try introducing the zero approximation (\ref{16}) for
$f_{s}^{r}(\beta-\beta_{1})$ into the second right-hand term of (\ref{14}).
Such a substitution means that we should further remain within the first
approximation in $n$ for the kernel of Eq. (\ref{14}). It follows, that in
this approximation equation (\ref{14}) becomes Markovian and can be formally
integrated. The result is%
\begin{equation}
f_{s}^{r}(\beta)=\exp[-U_{s}\beta-P_{s}U_{s,N-s}\int\limits_{0}^{\beta}%
d\beta^{^{/}}e^{-Q_{s}U_{s,N-s}\beta^{/}}],\label{16a}%
\end{equation}
where we have used that $f_{s}^{r}(0)=1$.

In order to calculate the integral in (\ref{16a}), we consider the expansion
(\ref{15}) of $e^{-Q_{s}U_{s,N-s}\beta^{/}}$ in $P_{s}U_{s,N-s}$. Remaining in
the first order in $P_{s}U_{s,N-s}$, we substitute $e^{-Q_{s}U_{s,N-s}%
\beta^{/}}$ in (\ref{16a}) with the first term in the right-hand side of
(\ref{15}). Thus, in the considered approximation we obtain%
\begin{equation}
f_{s}^{r}(\beta)=\exp[-U_{s}\beta+P_{s}e^{-U_{s,N-s}\beta}-1].\label{17}%
\end{equation}

In the adopted first in the density approximation for the kernel $-U_{s}%
\beta+P_{s}e^{-U_{s,N-s}\beta}-1$, the projected exponential $P_{s}%
e^{-U_{s,N-s}\beta}$ in (\ref{17}) should be calculated in the first order in
$n$ as to the terms containing energy of interaction of the $s$ - cluster
particles with remaining $N-s$ particles, i.e. $\Phi_{ij}$ ($1\leq i\leq
s,s+1\leq j\leq N$), but all terms related to the mutual interactions of the
irrelevant $N-s$ particles should be calculated in all orders in the
corresponding $\Phi_{ij}$ (see (\ref{16'''}) and the subsequent comment). It
can be easily done, if we take into account that the quantity $P_{s}%
e^{-U_{s,N-s}\beta}$ can be expressed in terms of the canonical $s$ - particle
distribution function $F_{s}(\beta)$ (\ref{6a}), related to the relevant
distribution function $f_{s}^{r}(\beta)$ by Eq. (\ref{6}), as%
\begin{equation}
P_{s}e^{-U_{s,N-s}\beta}=\frac{Z_{N}(\beta)}{V^{N}}e^{U_{s}\beta}F_{s}%
(\beta).\label{18}%
\end{equation}

We can now use the known density expansion of $F_{s}(\beta)$%
\begin{equation}
F_{s}(\beta;\{x_{i}\}_{s})=e^{-U_{s}\beta}[1+\sum\limits_{k=1}^{\infty}%
n^{k}\beta_{k}^{/}(\beta;\{x_{i}\}_{s})],\label{19}%
\end{equation}
where the factors $\beta_{k}^{/}(\beta;x_{1},\ldots,x_{s})$ are the cluster
integrals corresponding to cluster diagrams with the fixed $s$ points $x_{i}$
($1\leq i\leq s$) of the Ursell-Mayer theory (for details see, e.g.,
\cite{Salpeter 1958}). In particular, for a pair distribution function
$F_{2}(\beta;x_{1},x_{2})$ we should make in (\ref{19}) the substitutions
$U_{s}\rightarrow\Phi_{12}$, $\beta_{1}^{/}(x_{1},\ldots,x_{s})\rightarrow
\beta_{1}^{^{\prime}}(x_{1},x_{2})$, where
\begin{equation}
\beta_{1}^{\prime}(\beta;x_{1},x_{2})=\int dx_{3}f_{13}f_{23}.\label{19a}%
\end{equation}
\qquad According to the adopted approximation for the kernel of Eq.
(\ref{14}), we should insert the linear in $n$ term of (\ref{19}) into
(\ref{18}). Thus, in the considered approximation we obtain from (\ref{17}) -
(\ref{19}) and (\ref{6})
\begin{equation}
F_{s}(\beta;\{x_{i}\}_{s})=\frac{V^{N}}{Z_{N}(\beta)}\exp[\frac{Z_{N}(\beta
)}{V^{N}}-1]e^{-U_{s}\beta}\exp\left[  \frac{Z_{N}(\beta)}{V^{N}}n\beta
_{1}^{/}(\beta;\{x_{i}\}_{s})\right]  .\label{20}%
\end{equation}
where the first two factors depend only on the temperature, and the subsequent
exponentials define the dependence of the distribution function on the
particles coordinates.

Now we can use the following known density expansion for the partition
function $Z_{N}(\beta)$ (see, e.g., \cite{Isihara 1971})%
\begin{equation}
\frac{Z_{N}(\beta)}{V^{N}}=\exp[N\sum\limits_{k=1}^{\infty}n^{k}\frac
{\beta_{k}(\beta)}{k+1}],\label{20a}%
\end{equation}
where $\beta_{k}(\beta)$ are Mayer's irreducible cluster integrals. For
example the first irreducible cluster integral is defined as%
\begin{equation}
\beta_{1}(\beta)=\int[e^{-\beta\Phi(r)}-1]d\mathbf{r.}\label{20b}%
\end{equation}
Note, that the formal expansion of the partition function $Z_{N}(\beta)$
(\ref{20a}) in the density $n=N/V$%
\begin{equation}
Z_{N}(\beta)=V^{N}[1+N\sum\limits_{k=1}^{\infty}n^{k}\frac{\beta_{k}(\beta
)}{k+1}+\ldots]\label{20c}%
\end{equation}
leads to the correct equation of state virial expansion, i.e. $Nn^{k}\beta
_{k}/(k+1)$ can formally be regarded as a small parameter $\thicksim n^{k}$
\cite{Landau}.

In the spirit of the adopted approximation we retain only the first term of
$Z_{N}(\beta)/V^{N}-1$ expansion (\ref{20c}) (containing linear in $n$ term)
into the first exponential of (\ref{20}), and take $Z_{N}(\beta)/V^{N}=1$ in
the third exponential. Thus, we finally obtain from (\ref{20}) and (\ref{20a}
)
\begin{equation}
F_{s}(\beta;x_{1},\ldots,x_{s})=e^{-U_{s}\beta}\exp[n\beta_{1}^{/}(\beta
;x_{1},\ldots,x_{s})].\label{21}%
\end{equation}
Expansion of (\ref{21}) in the particle density agrees, naturally, with
(\ref{19}).

The solution for the distribution function (\ref{21}) going beyond the result
(\ref{19}) (in the linear in $n$ approximation, which can turn insufficient)
and takes into account the infinite in the density series of terms (diagrams).
This demonstrates the advantage of the expansion of the kernel (of the
equation for a function) as compared to the corresponding expansion of the
function itself. We can expect that the result (\ref{21}) will provide a
reasonable approximation for not very large densities.

One can try to improve Eq. (\ref{21}) by adding the higher order in $n$ terms
of the expansion (\ref{19}) to the kernel (\ref{18}). For example, by taking
into account the quadratic in $n$ term of (\ref{19}) we obtain%
\begin{equation}
F_{s}(\beta;\{x_{i}\}_{s})=e^{-U_{s}\beta}\exp[n\beta_{1}^{/}(\beta
;\{x_{i}\}_{s})+n^{2}\beta_{2}^{/}(\beta;\{x_{i}\}_{s})].\label{22}%
\end{equation}
The approximation (\ref{22}) (although not quite consistent with the accepted
approach to the kernel expansion) accounts for much more Mayer's diagrams than
the linear approximation for the kernel (\ref{21}).

Next terms of expansion (\ref{15}) contribute to the higher powers in the
projector $P_{s}$ terms of Eq. (\ref{14}) kernel expansion and may change the
obtained results.

Generally, the non-Markovian equation (\ref{13}) or (\ref{14}) is to be solved
with some approximation for the kernel. This may be done, e.g., with the help
of the Laplace transform of Eq. (\ref{13}).

\section{Conclusion}

We have obtained the exact closed equation (\ref{13}) (or (\ref{14})) for the
$s$ - particle distribution function which is (from our point of view)
surprising and of principal importance, because the known closed equations,
such as PY and HNC equations or those which follow from disentangling the
BBGKY chain, are approximate equations. It is not also quite clear how to
improve the mentioned approximate equations in the regular way by taking into
account the additional terms of, e.g., the expansion in the particle density.
Our equation is an integra-differential equation with respect to the inverse
temperature $\beta$ variable (the mentioned known equations are the nonlinear
equations with respect to the coordinate variables). The projection operator
approach to the Bloch equation for the classical distribution function of the
system of $N\gg1$ particles, which enabled us to reach a goal, is, as to our
knowledge, a new one. To some extent this method is reminiscent of the
approach used in the paper \cite{Montroll} (see also \cite{Isihara 1971}),
where the solution of the Bloch equation and the partition function for the
system of quantum particles are represented as the series in the particles'
interaction in the $\{x_{i},\beta\}$ space. For obtained rather complicated
non-Markovian ($\beta$ - convolution) equation the methods of its solution
should be developed. One of the possibilities is to expand the kernel of this
equation into the particle density series. The expansion of the kernel of
equation for the distribution function is generally much more effective than
that of this distribution function which is characteristic of the Ursell -
Mayer theory. That is why, the finding of the closed equations for reduced
distribution functions is important for the theory of dense gases and liquids.
We have suggested the expansion of the obtained equation kernel in the
particle density and found the solution for the $s$ - particle distribution
function in the linear in $n $ approximation for the kernel.

\

\end{document}